%% file: main.tex
\documentclass{article}
\usepackage[utf8]{inputenc}

\pdfoutput=1
\usepackage{import}
\usepackage{amsmath}
\usepackage{amsthm}
\usepackage{xspace}
\usepackage{comment}
\usepackage{natbib}
\usepackage{url}
\usepackage{enumitem}
\usepackage{hyperref}

\usepackage[capitalise]{cleveref}
\usepackage{tikz}

\usepackage{algorithm}
\usepackage{algorithmic}
\usetikzlibrary{shapes,arrows,positioning,fit,arrows.meta}

\input{macros}

\usepackage[margin=1in]{geometry}
\title{Full Proportional Justified Representation}
\author{Yusuf Hakan Kalayci\thanks{University of Southern California, kalayci@usc.edu} \and Jiasen Liu\thanks{University of Southern California, jliu3566@usc.edu} \and David Kempe\thanks{University of Southern California, David.M.Kempe@Gmail.com}}

\date{}
\begin{document}

\maketitle

\input{abstract}

\input{introduction}

\input{preliminaries}

\input{fpjr}

\input{hardness}

\input{conclusion}

\bibliographystyle{plainnat}
\bibliography{../davids-bibliography/names,../davids-bibliography/conferences,../davids-bibliography/publications,../davids-bibliography/bibliography,../main}

\newpage

\appendix

\input{appendix}

\end{document}

%% file: macros.tex
\newtheorem{theorem}{Theorem}[section]
\newtheorem{corollary}[theorem]{Corollary}
\newtheorem{lemma}[theorem]{Lemma}
\newtheorem{proposition}[theorem]{Proposition}

\newtheorem{definition}[theorem]{Definition}

\newtheorem{example}[theorem]{Example}

\newcommand{\set}[1]{\left\{ #1 \right\}}
\newcommand{\Set}[2]{\set{#1 \mid #2}}

\providecommand{\Kth}[1]{\ensuremath{{#1}^{\rm th}}}

\def\A{\mathcal{A}}

\newcommand{\eat}[1]{}

\newenvironment{lp*}{\begin{equation*}  \begin{array}{lll}}{\end{array}\end{equation*}}

\providecommand{\Kth}[1]{\ensuremath{{#1}^{\rm th}}}
\providecommand{\Ceiling}[1]{\ensuremath{\lceil {#1} \rceil}\xspace}
\providecommand{\Floor}[1]{\ensuremath{\lfloor {#1} \rfloor}\xspace}
\providecommand{\SET}[1]{\ensuremath{\{ #1 \}}\xspace}
\providecommand{\Set}[2]{\ensuremath{\SET{#1 \mid #2}}\xspace}



%% file: abstract.tex
\begin{abstract}
    In multiwinner approval voting, selecting a proportionally representative committee based on the voters' approval ballots is an essential task. The notion of justified representation (JR) 
    demands that any large ``cohesive'' group of voters should be proportionally ``represented''. Different specific definitions of justified representation define ``cohesiveness'' in different ways; two common ways are the following: \textbf{(C1)} the coalition unanimously approves a subset of candidates whose size is proportional to its share of the electorate, and \textbf{(C2)} each voter in the coalition approves at least a fixed fraction of a candidate subset proportional to the coalition's size. Similarly, among others, the following two concrete definitions of ``representation'' have been considered: \textbf{(R1)} the coalition's  collective utility from the winning set exceeds that of any proportionally sized alternative, and \textbf{(R2)} for any proportionally sized alternative, at least one member of the coalition derives less utility from it than from the winning set.
    
    Three of the four possible combinations have been extensively studied and used to define extensions of Justified Representation:
    \begin{itemize}
        \item \textbf{(C1)-(R1)}: Proportional Justified Representation (PJR)
        \item \textbf{(C1)-(R2)}: Extended Justified Representation (EJR)
        \item \textbf{(C2)-(R2)}: Full Justified Representation (FJR)
    \end{itemize}
    All three have merits, but also drawbacks. PJR is the weakest notion, and perhaps not sufficiently demanding; EJR may not be compatible with perfect representation; and it is open whether a committee satisfying FJR can be found efficiently.
    
    We study the combination \textbf{(C2)-(R1)}, which we call \emph{Full Proportional Justified Representation (FPJR)}. We investigate FPJR's properties and find that it shares advantages with PJR over EJR; specifically, several desirable proportionality axioms --- such as priceability and perfect representation --- imply FPJR and PJR but not EJR. 
    Next, we show that efficient rules like the greedy Monroe rule and the method of equal shares satisfy FPJR, thus matching one of the key advantages of EJR over FJR. 
    However, the Proportional Approval Voting (PAV) rule may violate FPJR, so neither of EJR and FPJR implies the other.
\end{abstract}

%% file: introduction.tex
\section{Introduction}

Selecting representatives from a large set is a fundamental problem with widespread applications, including political elections and selection of committees or advisory bodies \citep{ebadian:micha:boosting-sortition, faliszewski:skowron:multiwinner, caragiannis:shah:metric-distortion,pierczynski:skowron:approval}, selection of projects and participatory budgeting \citep{aziz:lee:talmon:pb,peters:grzegorz:mes-pb,peters:pierczynski:skowron:fjr}, and selection of representative documents or training sets in machine learning \citep{sanchez:fisteus:approval-data}.
A commonly accepted principle for representation is proportionality \citep{dummett-voting-procedures, humphreys:proportional-representation, moulin:choosing-tournament}: subgroups of the population should be represented in the selected set proportionally to their size.
In other words, if a cohesive subset $S$ constitutes a $\theta$ fraction of the population, then approximately a $\theta$ fraction of the representative set should reflect the preferences of $S$.
Naturally, many different concrete instantiations of this principle are possible, depending on how \emph{cohesiveness} and \emph{representation} are defined, and what type of information voters communicate about their preferences.

A common framework in practice, and the focus of our work, is \emph{approval-based multi-winner voting}.
In this setting, the $n$ voters submit approval ballots, listing all candidates they approve of.
The voting rule needs to choose a committee of given size $k$.

The study of voting rules achieving some sense of proportionality in this setting dates back well over a century, starting with the work in the 1890s of Thiele and Phragm\'{e}n, who sought proportional representation for minorities in parliament \citep{janson:phragmen-thiele}.
The fundamental approach of Thiele is to maximize total voter satisfaction, leading to rules such as Proportional Approval Voting and the Chamberlin-Courant rule \citep{thiele:pav, chamberlin:courant}.
Phragm\'{e}n's approach is to balance representation among voters, leading to rules such as seq-Phragm\'{e}n and leximax-Phragm\'{e}n.
A somewhat similar objective is pursued by \citet{monroe:rule}, whose rule assigns each voter to one representative, choosing candidates to ensure that each representative is supported by an equal-sized group of voters.

The proliferation of plausible committee selection rules makes it necessary to compare the guarantees their outcomes provide.
A very common and successful approach to this goal in social choice is to define \emph{axioms} which the voting rules are supposed to satisfy \citep{brandt:conitzer:handbook,lackner:skowron:abc-survey}. 
Following this approach, \citet{aziz:brill:justified} were the first to introduce axioms for proportional representation under approval ballots.
They proposed the notion of \emph{Justified Representation} (JR) by adapting the concept of core stability from cooperative game theory.
Theirs was the first step toward constructing a broader family of axioms to assess how fair or proportional a committee is.
JR considers a coalition as ``cohesive'' if it constitutes a $1/k$ fraction of the population and unanimously agrees on a single candidate.
A coalition is deemed ``represented'' if at least one member has positive utility --- meaning that they approve at least one winner.
A committee satisfies JR if every cohesive coalition is represented in this way.

This formulation of ``cohesiveness'' and ``representation'' has subsequently been extended to formulate more demanding proportionality axioms, reviewed in more detail in Section~\ref{sec:axioms}.
Specifically, \emph{Proportional Justified Representation} (PJR) \citep{sanchez:elkind:pjr} defines a coalition as $\ell$-cohesive if it constitutes an $\frac{\ell}{k}$ fraction of the population and unanimously agrees on $\ell$ candidates.
An $\ell$-cohesive coalition is considered ``represented'' if the coalition's collective utility is at least $\ell$, i.e., the union of approval sets includes at least $\ell$ winners.
\emph{Extended Justified Representation} (EJR) \citep{aziz:brill:justified} makes the ``representation'' condition more stringent by requiring that there exists a voter within the coalition with utility $\ell$. 

An even more demanding axiom, \emph{Full Justified Representation} (FJR) \citep{peters:pierczynski:skowron:fjr}, weakens the notion of ``cohesiveness'' and requires more coalitions to be represented.
A coalition is considered \emph{weakly $\ell$-cohesive} if, for some witness set of candidates (with size proportional to the coalition size), each voter in the coalition approves at least $\ell$ candidates from this set.
FJR requires that such coalitions be represented according to the EJR criterion.

These axioms can be visualized along two dimensions: the notion of cohesiveness (unanimous agreement vs.~densely approved small witness set) and the notion of representation (collective utility vs.~maximum utility).
The axioms PJR, EJR, and FJR cover three out of four possible combinations of these notions.
In our work, we explore the fourth combination, which merges the cohesiveness notion of densely approved witness set with the representation notion of collective utility.
We call this axiom \emph{Full Proportional Justified Representation} (FPJR), as it combines features of both Full Justified Representation and Proportional Justified Representation.
More formally (a precise definition is given in Section~\ref{sec:fpjr}), FPJR requires that the collective utility of every weakly $\ell$-cohesive coalition is at least~$\ell$.

Our goal in this work is to understand how FPJR relates to other proportionality axioms and related properties, and which algorithms guarantee that their outputs satisfy FPJR. 
Here, we overview some of the observed relationships.

PJR is the weakest axiom among the ones introduced above.
However, \citet{sanchez:elkind:pjr} showed that PJR is compatible with another proportionality notion called \emph{Perfect Representation} (PER).
A committee satisfies Perfect Representation if there exists an equal-sized $k$-partition of voters such that each part unanimously approves a distinct winner.
While a committee satisfying PER may violate EJR (and thus FJR), any committee with Perfect Representation satisfies PJR \citep{sanchez:elkind:pjr}.
Furthermore, \citet{peters:skowron:priceability} introduced another proportionality axiom, \emph{priceability}, which justifies committees as the result of voters spending equal amounts of money to elect candidates.
They showed that any priceable committee also satisfies PJR.

Our work reveals that FPJR, like EJR, is a stronger requirement than PJR. Nonetheless, paralleling PJR (and contrasting EJR), any committee meeting Perfect Representation or priceability also meets FPJR. This insight enriches our understanding of priceability, as we now see it can fulfill an even more stringent axiom than PJR.

An important property of a proportionality axiom is the existence of efficient (polynomial-time) algorithms to find a committee that satisfies it.
In the literature, several efficient rules are known to satisfy the EJR axiom (and hence PJR), including LS-PAV (a local search variant of PAV), the EJR-Exact rule, and the Method of Equal Shares \citep{brams:brill:excess-method, aziz:elkind:complexity}.
In contrast, other efficient procedures --- such as the Phragm\'{e}n-type rules (seq-Phragm\'{e}n and leximax-Phragm\'{e}n) and the greedy Monroe rule (for the special case $k \mid n$) --- satisfy PJR but not EJR \citep{faliszewski:slinko:greedy-monroe, brill:freeman:phragmen-JR}.
In this work, we show that the Monroe rule and its efficient greedy variant satisfy FPJR, as do the Method of Equal Shares and Phragm\'{e}n-type rules.
However, the PAV and LS-PAV rules may violate FPJR, so committees satisfying PJR or EJR may not satisfy FPJR.
Conversely, the Monroe rule violates EJR but satisfies FPJR, implying that FPJR and EJR are incomparable.

Although the FJR axiom is the most demanding among the family of axioms discussed, it remains an open question whether any efficient rule can output a committee satisfying FJR.
Moreover, it is known that all currently known efficient rules violate the FJR axiom.
FPJR shares the same cohesiveness condition with FJR, and we hope that the properties of FPJR explored in our work may shed further light on the algorithmic properties of FJR.

Finally, verifying whether a committee satisfies a certain axiom is also an important task.
\citet{aziz:brill:justified} showed that verifying EJR is coNP-complete, and \citet{aziz:elkind:complexity} showed the same for PJR.
In this work, we extend these results and show that verifying FJR, FPJR, and core stability is also coNP-complete.
Core stability is one of the strongest proportionality notions, ensuring that for any coalition and any alternative subset of candidates of size proportional to the size of the coalition, there exists a voter who weakly prefers the committee to the alternative set.\footnote{One of the big open questions in the area of approval-based committee selection is whether the core is always non-empty.}
Our hardness results for FJR, FPJR, and the core are based on the same reductions from the \textsc{Balanced Biclique} problem as the hardness results of \citet{aziz:brill:justified,aziz:elkind:complexity}, but with a more involved analysis.

In summary, our main contributions are the following:
\begin{enumerate}[label=(\roman*)]
    \item We introduce the \emph{Full Proportional Justified Representation} (FPJR) axiom, filling a gap in the family of proportionality axioms by combining the cohesiveness notion of densely approved witness set with the representation notion of collective utility.
    \item We demonstrate that FPJR is a stronger notion than PJR, yet any committee satisfying Perfect Representation or priceability also satisfies FPJR.
    \item We show that the Monroe rule and its efficient greedy variant, as well as the Method of Equal Shares and Phragm\'{e}n-type rules, satisfy FPJR.
      However, the PAV and LS-PAV rules may violate FPJR.
    \item We establish that verifying whether a committee satisfies FPJR, FJR, or core stability is coNP-complete, extending known results about the complexity of verifying proportionality axioms.
\end{enumerate}

\subsection*{Related Work}

In addition to the most immediately related work discussed previously, the following strands of literature relate to our present work.

\subsubsection*{Verifiable Justified Representation Axioms}

Recently, \citet{brill:peters:verifiable-JR} introduced proportionality axioms of the Justified Representation type called EJR+ and PJR+.
In a given election and committee, a coalition is called \emph{$\ell$-deprived} if it constitutes at least an $\frac{\ell}{k}$ fraction of the population and unanimously agrees on a non-winner candidate.
The PJR+ axiom requires that any $\ell$-deprived coalition must have a collective utility of at least $\ell$.
Similarly to EJR, EJR+ requires that for any $\ell$-deprived coalition, there exists a voter in the coalition with utility at least $\ell$.
\citet{brill:peters:verifiable-JR} showed that EJR+ is more demanding than EJR, and PJR+ is more demanding than PJR.
Importantly, and in contrast to PJR and EJR, both axioms can be verified in polynomial time.

The PAV rule, which satisfies EJR+ and PJR+, violates FPJR.
This implies that committees satisfying EJR+ or PJR+ might violate FPJR.
Conversely, the Monroe rule violates EJR and thus EJR+, yet it satisfies FPJR.
This implies that FPJR and EJR+ are incomparable.
Additionally, in Appendix~\ref{apx:fpjr_vs_pjr+}, we provide an example showing a committee satisfying FPJR but violating PJR+, further implying that FPJR and PJR+ are incomparable.

\subsubsection*{Participatory Budgeting}

Participatory Budgeting (PB) is a democratic process for deciding on the funding of public projects, adopted in several cities worldwide.
In PB, each election consists of a set of voters, candidates, a ballot profile, a budget, and a cost function describing the cost of each candidate.
The total cost of the candidates in the winning set must not exceed the budget.
Various types of ballot profiles have been studied for PB, including ranked ballots \citep{aziz:lee:proportional-pb-ordinal}, Knapsack ballots \citep{goel:krishnaswamy:knapsack-voting}, and approval ballots \citep{peters:grzegorz:mes-pb}.
For a detailed discussion of PB, we refer readers to the recent surveys by \citet{aziz:shah:pb} and by \citet{maly:rey:pbsurvey}.

One key goal in PB is again proportional representation \citep{brill:forster:abc-pb-proportionality}.
In the approval voting setting, existing proportionality axioms such as EJR \citep{peters:grzegorz:mes-pb}, PJR, FJR, EJR+, and PJR+ \citep{brill:peters:verifiable-JR} have been generalized to the PB context as well.
Moreover, methods like the Method of Equal Shares \citep{peters:grzegorz:mes-pb} are known to satisfy these axioms in the PB domain.
One can follow a similar approach to generalize the FPJR axiom to participatory budgeting.
However, this is beyond the focus of this paper, and we leave the investigation of FPJR for participatory budgeting as a direction for future work.

\subsubsection*{Proportional Representation Beyond Approval Voting}

The concept of proportional representation has also been studied beyond approval voting.
Early research by \citet{dummett-voting-procedures} aimed to ensure proportionality for \emph{Solid Coalitions} (PSC) on ranked ballots.
These coalitions consist of voter groups whose sets of top candidates are the same (for corresponding set sizes), though the ordering of these top candidates may differ within the coalition.
Another notion of proportional representation appears for ranked ballots compatible with a hidden distance function determined by a metric space \citep{kalayci:kempe:kher:proportional-representation-metric}.
In this setting, the cost of a candidate to a voter is measured by this hidden distance function, and the goal is to ensure that any sufficiently large coalition has an (approximately) smaller cost with the winning set than with a proportionally sized alternative.
Note that while the benchmark is a hidden distance function, the ballot profile consists of rankings only.

Finally, \citet{skowron:lackner:proportional-rankings} extended the principle of proportional representation to rankings: given approval preferences, the goal is to generate aggregated rankings so that cohesive groups of voters are represented proportionally in each initial segment of the ranking.

%% file: preliminaries.tex
\section{Preliminaries}
We consider an election with a set $V$ of $n$ voters, and a set $C$ of $m$ candidates.
Each voter $v \in V$ submits an \emph{approval ballot} $A_v \subseteq C$, listing the subset of candidates that $v$ approves.
The vector of all approval ballots is denoted by $\A = (A_v)_{v \in V}$ and referred to as the \emph{ballot profile}.
For any candidate $c \in C$, the set of voters approving $c$ is denoted by $N_c$.
For a desired committee size $k > 0$, an \emph{approval-based multi-winner voting rule} takes as input the election $(V, C, \A, k)$ and outputs a subset $W \subseteq C$ of size $k$, called the \emph{winning set} or \emph{committee}.

\subsection{Proportional Representation Axioms}
\label{sec:axioms}
Proportional representation axioms express that cohesive groups of voters should receive fair representation in the committee. Following the foundational work of \citet{aziz:brill:justified}, various definitions have been proposed to formalize the notions of ``cohesiveness'' and ``representation.''


\begin{definition} \label{def:cohesive}
    Consider an approval-based multi-winner election $(V, C, \A, k)$. 
    A coalition $S \subseteq V$ is called \emph{$\ell$-cohesive} if $\frac{|S|}{n} \geq \frac{\ell}{k}$ and $\left| \bigcap_{v \in S} A_v \right| \geq \ell$.
\end{definition}

That is, $S$ is $\ell$-cohesive if it comprises at least an $\frac{\ell}{k}$ fraction of the population (and is therefore ``entitled'' to at least $\ell$ out of the $k$ committee members), and the voters in $S$ agree on at least $\ell$ candidates.

A weaker form of cohesiveness can be defined by relaxing the condition of having $\ell$ common approved candidates.

\begin{definition} \label{def:weak-cohesive}
  A coalition $S \subseteq V$ is called \emph{weakly $\ell$-cohesive with witness set $T$} 
  if $T$ is a subset of candidates such that $\frac{|S|}{n} \geq \frac{|T|}{k}$, and $|A_v \cap T| \geq \ell$ for each voter $v \in S$.
  $S$ is called \emph{weakly $\ell$-cohesive} if there exists a witness $T$ such that $S$ is weakly $\ell$-cohesive with witness $T$.
\end{definition}

Weak cohesiveness expresses that $S$ is large enough to deserve a candidate set of the size of $T$, and each voter in $S$ approves at least $\ell$ candidates from $T$.

Based on these concepts of cohesiveness, the following proportionality axioms have been introduced: 

\begin{itemize}
    \item \textbf{PJR \citep{sanchez:elkind:pjr}:} A committee $W$ satisfies \emph{Proportional Justified Representation (PJR)} if, for every $\ell \geq 1$ and every $\ell$-cohesive coalition $S$, the voters in $S$ collectively approve at least $\ell$ candidates in $W$, i.e.,
    \[
    \left| W \cap \bigcup_{v \in S} A_v \right| \geq \ell.
    \]
    \item \textbf{EJR \citep{aziz:brill:justified}:} A committee $W$ satisfies \emph{Extended Justified Representation (EJR)}  if, for every $\ell \geq 1$ and every $\ell$-cohesive coalition $S$, there exists a voter $v \in S$  approving at least $\ell$ candidates in $W$, i.e., $|A_v \cap W| \geq \ell$.

  \item \textbf{FJR \citep{peters:pierczynski:skowron:fjr}:} A committee $W$ satisfies \emph{Full Justified Representation (FJR)} if, for every $\ell \geq 1$ and every weakly $\ell$-cohesive coalition $S$, there exists a voter $v \in S$ approving at least $\ell$ candidates in $W$, i.e., $|A_v \cap W| \geq \ell$.
  \item \textbf{Core Stability \citep{aziz:brill:justified}:} A committee $W$ is in the \emph{core} if for every coalition $S$ and subset of candidates $T$ such that $\frac{|S|}{n} \geq \frac{|T|}{k}$, there exists a voter $v \in S$ approving at least as many candidates in $W$ as in $T$, i.e., $|A_v\cap W| \geq |A_v\cap T|$.
\end{itemize}

It is well known \citep{aziz:brill:justified,sanchez:elkind:pjr,peters:pierczynski:skowron:fjr}
that core stability implies FJR, FJR implies EJR, and EJR implies PJR. 

Beyond the ``core'' framework and its Justified Representation variants, we consider two other notions of proportionality from the literature.
First,  a set of candidates $W$ provides \emph{perfect representation (PER)} \citep{sanchez:elkind:pjr} for $\A$ and $k$ if the voter set $V$ can be partitioned into $k$ equal-size pairwise disjoint subsets $V_1, \dots, V_k$, with the following property:
one can assign a distinct candidate $c_i$ from $W$ to each subset $V_i$ such that every voter in $V_i$ approves of {$c_i$}.
In particular, PER can only be satisfied when $k$ divides $n$.

Recently, \citet{peters:skowron:priceability} introduced the concept of \emph{priceability} for committees. Intuitively, a committee is \emph{priceable} if it can be justified as the result of voters spending equal amounts of money to elect candidates. Formally, a \emph{price system} is a pair $(p, (p_v)_{v \in V})$ comprising a price $p > 0$ and a \emph{payment function} $p_v : C \to [0,1]$ for each voter $v$. Each payment function $p_v$ satisfies that $p_v(c) > 0$ only for candidates $c \in A_v$ that the voter approves, and $\sum_{c \in C} p_v(c) \leq 1$, ensuring the voter spends at most one unit.

A price system \emph{supports} a committee $W$ if for each candidate $c \in W$, the payments sum to $p$ (i.e., $\sum_{v \in V} p_v(c) = p$), no  candidate $c \notin W$ receives any payment, and no unelected candidate has supporters whose remaining unspent budget strictly exceeds $p$, i.e., for any fixed $c \notin W$:
\[
\sum_{v\in N_{c}} \left( 1 - \sum_{c' \in W} p_v(c') \right) \leq p.
\]

A committee $W$ is \emph{priceable} if a supporting price system exists. If $W$ is supported by a price system with price $p$, then $p \leq \frac{|V|}{|W|}$ since the total voter spending is exactly $p \cdot |W|$, which cannot exceed the total budget $|V|$. Notably, priceability does not impose any constraints on the target committee size $k$.

We relate priceability and PER with the following immediate proposition, which states that PER is a more demanding axiom than priceability.

\begin{proposition} \label{prop:PER-implies-priceability}
Any committee providing perfect representation is also priceable.
\end{proposition}

To see why this proposition holds, recall that a perfect representation partitions voters into groups and associates each group with a distinct candidate whom the group unanimously approves.
In this setting, each group can spend their total budget on the associated candidate.

\subsection{Approval-based Multi-winner Rules}
Next, we review several approval-based multi-winner voting rules. For a detailed examination of these rules and their various axiomatic properties, we refer readers to the survey by \citet{lackner:skowron:abc-survey}.

\subsubsection{Proportional Approval Voting (PAV)}
The Proportional Approval Voting (PAV) rule \citep{thiele:pav} selects a committee $W$ maximizing the PAV score 
\[
\text{PAV}(W) = \sum_{v \in V} H \left( |A_v \cap W| \right),
\]
where $H(t) = \sum_{i=1}^{t} \frac{1}{i}$ is the \Kth{t} harmonic number, and $|A_v \cap W|$ is the number of candidates from $W$ approved by voter $v$.
The PAV rule satisfies EJR and thus PJR \citep{aziz:brill:justified}.
Although computing the exact PAV outcome is NP-hard \citep{aziz:gaspers:computational}, there exists a polynomial-time local search variant, called LS-PAV, that produces a committee with an approximately optimal PAV score and still satisfies EJR \citep{aziz:elkind:complexity}.

\subsubsection{Monroe's Rule}
Monroe's Rule \citep{monroe:rule} aims to achieve proportional representation by matching voters to candidates in the committee, achieving an approximation to perfect representation.
Specifically, it seeks a committee $W$ and an assignment $\pi: V \rightarrow W$ such that:
\begin{enumerate}[label=(\roman*)]
\item Each candidate $c \in W$ is assigned approximately $\frac{n}{k}$ voters:
  $\Floor{\frac{n}{k}} \leq |\pi^{-1}(c)| \leq \Ceiling{\frac{n}{k}}$.
    \item Defining the function $d(v, c) = 1$ if $c \in A_v$ and $0$ otherwise, the total satisfaction $\sum_{v \in V} d(v, \pi(v))$ is maximized among all possible assignments.
\end{enumerate}
In particular, if $\sum_v d(v, \pi(v)) = n$ and $k \mid n$, then the resulting committee $W$ satisfies perfect representation. 
The Monroe Rule, although known to violate EJR, does satisfy PJR provided $k \mid n$. This divisibility condition is essential, as \citet{aziz:brill:justified} presented a counterexample when $k \not \mid n$.
Computing the exact Monroe outcome is NP-hard.
However, \citet{faliszewski:slinko:greedy-monroe}  developed a polynomial-time implementable greedy version of Monroe's Rule which also satisfies PJR when $k\mid n$.
This variant builds the committee iteratively, repeating the following while $|W| < k$: 
\begin{enumerate}[label=(\roman*)]
    \item Select a candidate $c \notin W$ who is approved by the largest number of unassigned voters.
    \item Assign approximately $\frac{n}{k}$ such voters to $c$, ensuring that each candidate in $W$ is assigned between $\Floor{\frac{n}{k}}$ and $\Ceiling{\frac{n}{k}}$ voters.
      Specifically, until the number of remaining voters is divisible by the number of remaining slots, it assigns $\Ceiling{\frac{n}{k}}$ voters; and subsequently assigns $\Floor{\frac{n}{k}}$. In each iteration, it adds as many voters as possible approving $c$; if there are not enough voters for a step, it selects some other arbitrary unassigned voters.
    \item Add $c$ to $W$.
\end{enumerate}

\subsubsection{Priceable Rules}
Within the landscape of approval voting rules, \emph{priceable rules} are guaranteed to output priceable committees. Two notable rules in this family are the Method of Equal Shares and Phragm\'{e}n's method \citep{peters:skowron:priceability}.

The Method of Equal Shares (ES) \citep{peters:skowron:mes,peters:grzegorz:mes-pb} constructs the committee $W$ sequentially. 
In each round, it checks which candidates can be ``purchased'' at cost $p = \frac{n}{k}$ by the voters who approve them, using their remaining budgets.
(Each voter starts with a total budget of $1$.)
Specifically, ES looks for a cost-sharing threshold $q$ such that every approving voter pays at most $q$, and these contributions collectively cover $p$. Among all candidates that can be afforded this way, the rule selects the candidate requiring the smallest $q$. Once selected, the voters’ budgets are reduced by their actual contributions. The process repeats until no additional candidate can be afforded, at which point the current committee $W$ is returned.

Phragm\'{e}n's methods ~\citep{janson:phragmen-thiele, brill:freeman:phragmen-JR}  select a committee by assigning \emph{loads} to voters who approve the elected candidates, aiming to balance these loads as evenly as possible. Each elected candidate's load is distributed among their approving voters, summing to $1$.
The objective is to minimize the maximum voter load, with prominent variations including leximax-Phragm\'{e}n and sequential Phragm\'{e}n.

While the Monroe rule outputs a perfect representation (and therefore a priceable committee) when a perfect representation exists \citep{brill:freeman:phragmen-JR}, it is still unclear if the Monroe rule outputs a priceable committee in general. In this paper, we present an example demonstrating that the Monroe rule can fail to satisfy priceability, even when the desired divisibility conditions are met.

%% file: fpjr.tex
\section{Full Proportional Justified Representation}
\label{sec:fpjr}

This section is devoted to our main results.
We begin by defining \emph{Full Proportional Justified Representation}.

\begin{definition}[Full Proportional Justified Representation (FPJR)] \label{def:fpjr}
  A committee $W$ satisfies Full Proportional Justified Representation (FPJR) if, for every weakly $\ell$-cohesive coalition $S$, we have $|W \cap \bigcup_{v \in S} A_v| \geq \ell$.
\end{definition}

We prove a useful property of weakly $\ell$-cohesive coalitions.

\begin{lemma}
    \label{lem:weak_cohesive_implies_cohesive}
    For any weakly $\ell$-cohesive coalition $S$ with witness $T$, there exists a candidate $c \in T$ such that $|N_c \cap S| \geq \ell \cdot \dfrac{n}{k}$.
\end{lemma}

\begin{proof}
  By definition of being weakly $\ell$-cohesive, we know that $\frac{|S|}{n} \geq \frac{|T|}{k}$,
  and $|A_v \cap T| \geq \ell$ for all $v \in S$.
  Thus, 
    \[
    \sum_{v \in S} |A_v \cap T| \geq \ell \cdot |S| \geq \ell \cdot |T| \cdot \dfrac{n}{k}.
    \]
    Hence, the average approval of candidates $c \in T$ is
    \[
      \frac{1}{|T|} \sum_{c \in T} |N_c \cap S|
      = \frac{1}{|T|} \sum_{v \in S} |A_v \cap T|  
      \geq \ell \cdot \dfrac{n}{k}.
    \]
    Because the average approval of candidates in $T$ is at least $\ell \cdot \dfrac{n}{k}$, there must exist at least one candidate $c \in T$ such that $|N_c \cap S| \geq \ell \cdot \dfrac{n}{k}$.
  \end{proof}

\input{fpjr_axiomatic}

\input{fpjr_monroe}

%% file: fpjr_axiomatic.tex
\subsection{Axiomatic Properties}
We begin by relating FPJR to previous proportionality axioms.
First, observe that FPJR requires any weakly $\ell$-cohesive coalition to collectively approve at least $\ell$ candidates from the committee $W$.
In contrast, FJR demands that there exist a voter $v$ within the coalition who \emph{individually} approves $\ell$ candidates from the committee.
This immediately implies that FJR is a more stringent axiom than FPJR. 
Indeed, we show below that the Monroe rule and the Method of Equal Shares satisfy FPJR, but may violate FJR.
This implies that FJR is a strictly more demanding axiom than FPJR \citep{peters:grzegorz:mes-pb}.

\begin{corollary} \label{cor:FJR-more-demanding-then-FPJR}
    FJR is a strictly more demanding axiom than FPJR.
\end{corollary}

Next, we investigate how FPJR relates to EJR and PJR.
First, we revisit an example from the literature \citep{peters:pierczynski:skowron:fjr,peters:skowron:priceability} illustrating that the Proportional Approval Voting (PAV) rule may violate the FJR axiom.

\begin{example}[PAV might violate FPJR \citep{peters:pierczynski:skowron:fjr,peters:skowron:priceability}] \label{ex:PAV-violates-FPJR} 
    Consider an election with $15$ candidates and $n = 6$ voters $V=\{v_1,v_2,v_3,v_4,v_5,v_6\}$, whose approval sets are as follows:
    \begin{align*}
        &A_{v_1}=\{c_1, c_2, c_3, c_4\} &A_{v_2}&=\{c_1, c_2, c_3, c_5\} &\\
        &A_{v_3}=\{c_1, c_2, c_3, c_6\} &A_{v_4}&=\{c_7, c_8, c_9\} &\\
        &A_{v_5}=\{c_{10}, c_{11}, c_{12}\} &A_{v_6}&=\{c_{13}, c_{14}, c_{15}\} &
    \end{align*}
    The committee size is $k = 12$. The PAV rule selects the committee $W = \{c_1, c_2, c_3, c_7, c_8, c_9, c_{10}, c_{11}, c_{12}, c_{13}, c_{14}, c_{15}\}$. Consider the weakly $4$-cohesive coalition $\SET{v_1, v_2, v_3}$  (with witness set of candidates $\SET{c_1, c_2, c_3, c_4, c_5, c_6}$). They collectively approve only $3$ candidates in the committee $W$. Therefore, FPJR is violated. Sequential PAV also selects this committee; hence, it may violate FPJR too.
\end{example}
Recall that the PAV rule satisfies EJR and, consequently, PJR. 
    On the other hand, by focusing on weakly $\ell$-cohesive sets with witness set size $|T| = \ell$, the requirements of FPJR coincide with those needed to satisfy PJR.
This implies that FPJR is a strictly more demanding axiom than PJR.

\begin{corollary} \label{cor:FPJR-more-demanding-than-PJR}
    FPJR is a strictly more demanding axiom than PJR. 
\end{corollary}

Next, we show that priceability is more demanding than FPJR.

\begin{theorem} \label{thm:priceability-implies-FPJR}
    Every priceable committee $W$ satisfies FPJR with $k=|W|$.
\end{theorem}

\begin{proof}
  Let $(p, (p_v)_{v \in V})$ be a price system for the committee $W$.
  For each voter $v$, let $b_v = 1 - \sum_{c \in W} p_v(c)$ be $v$'s remaining budget.

  Consider a weakly $\ell$-cohesive coalition $S$ with witness set $T$, and let $W_S$ denote the subset of committee members approved by at least one voter in $S$, i.e., $W_S = W \cap \bigcup_{v \in S} A_v$.
  Let $O = T \cap W_S$ be the subset of $W_S$ that is also in $T$.
  
  We lower-bound the sum, over all candidates $c \in T\backslash O$,
  of the remaining budget of voters approving $c$.
  Notice that voters approving multiple candidates in $T \setminus O$ will be counted multiple times in this sum. Because each voter $v \in S$ approves of at least $\ell$ candidates in $T$, of whom at most $|O|$ can be in $O$, we obtain that each $v \in S$ is in $N_c$ for at least $\ell - |O|$ candidates in $T \setminus O$.
  This gives us the bound
  
\begin{align}
  \sum_{c \in T \setminus O} \sum_{v \in N_c} b_v
        & \geq (\ell-|O|) \cdot \sum_{v \in S} b_v \nonumber\\
        & = (\ell - |O|) \cdot \sum_{v \in S} \left(1 - \sum_{c \in C} p_v(c)\right) \nonumber\\
        & \geq (\ell - |O|) \cdot \left( |S|-|W_S| \cdot p \right) \nonumber\\
        & \geq (\ell - |O|) \cdot \left( |T| \cdot \frac{n}{k} - |W_S| \cdot \frac{n}{k} \right); \label{eqn:budget-lower-bound}
\end{align}
here, the last inequality used the size bound on $S$ in the definition of cohesiveness and the fact that the price of each candidate must be bounded by $\frac{n}{k}$ to be feasible.
Furthermore, the last inequality is strict if $p < \frac{n}{k}$.

On the other hand, because none of the candidates $c \in T \setminus O$ were included in $W$, the remaining budget of the supporters for each such $c$ must add up to at most $p$, implying that
\begin{align}
  \sum_{c \in T \setminus O} \sum_{v \in N_c} b_v
  & \leq |T \setminus O| \cdot p
  \leq |T \setminus O| \cdot \frac{n}{k}. \label{eqn:budget-upper-bound}
\end{align}
Combining the two inequalities and canceling out the common term $\frac{n}{k}$, we obtain that
$|T| - |O| = |T \setminus O| \geq (\ell - |O|) \cdot (|T| - |W_S|)$.
We now consider two cases:
\begin{enumerate}
\item If $|W_S| \geq \ell$, then by definition, the representation condition is satisfied for $S$.
\item If $|T| = \ell$, then we can invoke the result from Proposition~1 in \citep{peters:skowron:priceability}, which states that priceability implies PJR,
to conclude that $|W_S| \geq \ell$, meaning that the representation condition is satisfied for $S$.
\end{enumerate}

Outside of these two cases, we have that $|T| > \ell > |W_S| \geq |O|$.
When $p < \frac{n}{k}$, Inequality~\eqref{eqn:budget-lower-bound} becomes strict, and the inequality
$|T| - |O| > (\ell - |O|) \cdot (|T| - |W_S|)$ must be strict also. However, because $(\ell-|O|) + (|T| - |W_S|) \geq |T| - |O| + 1$, and both factors $\ell-|O|$ and $|T|-|W_S|$ are strictly positive, this is impossible. 
    
Finally, we consider the case $p=\frac{n}{k}$.
Here, we note that the winners in $W \setminus W_S$ are not approved by any voters in $S$, and must therefore be entirely paid for by voters in $V \setminus S$.
Because $|S| \geq |T| \cdot \frac{n}{k} \geq \ell \cdot \frac{n}{k}$, there are at most $|V \setminus S| \leq (k-\ell) \cdot \frac{n}{k}$ such voters.
With their combined budget of at most $(k - \ell) \cdot \frac{n}{k}$, even if they do not contribute towards candidates in $W_S$, at a candidate price of $\frac{n}{k}$, they can support at most $(k-\ell)$ candidates.
But because $W_S$ contains strictly fewer than $\ell$ candidates, this leaves $W$ with strictly fewer than $k$ candidates, a contradiction.
\end{proof}
  
This theorem leads to several corollaries, which we summarize below.

Since a priceable committee may violate EJR~\citep{peters:skowron:priceability}, Example~\ref{ex:PAV-violates-FPJR} and Theorem~\ref{thm:priceability-implies-FPJR} imply that EJR and FPJR are incomparable.

\begin{corollary} \label{cor:EJR-FPJR-incomparable}
    EJR and FPJR are incomparable. 
\end{corollary}

As shown in Proposition~\ref{prop:PER-implies-priceability}, any committee satisfying Perfect Representation is also priceable and hence satisfies FPJR.
\begin{corollary}
    Perfect Representation implies FPJR.
\end{corollary}

Finally, \citet{peters:skowron:priceability} showed that the Method of Equal Shares and Phragm\'{e}n's rule always output priceable committees; hence, their outputs satisfy FPJR.

\begin{corollary} \label{cor:phragmen-and-ES-satisfy-FPJR}
    Phragm\'{e}n's rule and Equal Shares satisfy FPJR.
\end{corollary}

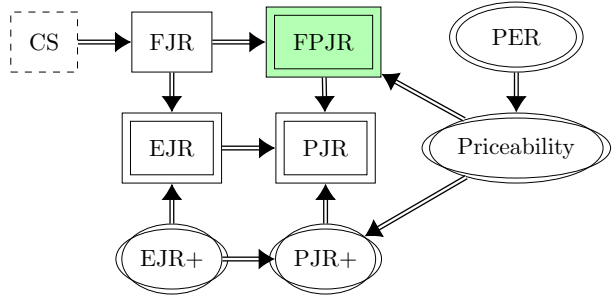
\begin{figure}[h] \centering 

    \resizebox{.49\textwidth}{!}{%
        \begin{tikzpicture}[
            scale=0.5,
            node distance = 0.6cm and 0.8cm,
            box/.style = {rectangle, draw, minimum width=1cm, minimum height=0.6cm},
            oval/.style = {ellipse, draw, minimum width=1.5cm, minimum height=0.8cm},
            arrow/.style = {double, double distance=1pt, line width=0.2mm, -{Latex[scale=1, width=9pt, length=7pt]} 
            },
            inner box/.style = {rectangle, draw, minimum width=0.4cm, minimum height=0.2cm, inner sep=0pt}
        ]
        
        \node[draw,dashed,minimum width=1cm, minimum height=0.9cm] (CS) {CS};
        \node[box, right=of CS, minimum width=1.2cm, minimum height=0.9cm] (FJR) {FJR};
        \node[box, right=of FJR, fill=green!30] (FPJR) {
            \begin{tikzpicture}
                \node[inner box, inner sep=8pt, minimum width=1cm, minimum height=5mm] {FPJR};
            \end{tikzpicture}
        };;

        \node[box, below=of FJR] (EJR) {
            \begin{tikzpicture}
                \node[inner box, inner sep=8pt, minimum width=1cm, minimum height=5mm] {EJR};
            \end{tikzpicture}
        };
        
        \node[box, right=of EJR] (PJR) {
            \begin{tikzpicture}
                \node[inner box, inner sep=8pt, minimum width=1cm, minimum height=5mm] {PJR};
            \end{tikzpicture}
        };
        \node[draw, ellipse, minimum width=1.5cm, minimum height=1.05cm, below=of EJR] (EJRp) {};
        \node[draw, ellipse, minimum width=1.3cm, minimum height=0.9cm] at (EJRp.center) {EJR+};
        
        \node[draw, ellipse, minimum width=1.5cm, minimum height=1.05cm, below=of PJR] (PJRp) {};
        \node[draw, ellipse, minimum width=1.3cm, minimum height=0.9cm] at (PJRp.center) {PJR+};
        
        \node[draw, ellipse, minimum width=2.6cm, minimum height=1.05cm, right=of PJR] (Price) {};
        \node[draw, ellipse, minimum width=2.4cm, minimum height=0.9cm] at (Price.center) {Priceability};

        \node[draw, ellipse, minimum width=2cm, minimum height=1.05cm, above=of Price] (PER) {};
        \node[draw, ellipse, minimum width=1.8cm, minimum height=0.9cm] at (PER.center) {PER};
        
        \draw[arrow] (CS) -- (FJR);
        \draw[arrow] (FJR) -- (FPJR);
        \draw[arrow] (FPJR) -- (PJR);
        \draw[arrow] (PER) -- (Price);
        \draw[arrow] (FJR) -- (EJR);
        \draw[arrow] (EJR) -- (PJR);
        \draw[arrow] (EJRp) -- (EJR);
        \draw[arrow] (EJRp) -- (PJRp);
        \draw[arrow] (PJRp) -- (PJR);
        \draw[arrow] (Price) -- (PJRp);
        \draw[arrow] (Price) -- (FPJR);
        
        \end{tikzpicture}
    }
    \caption{
        In the diagram, we illustrate how proportionality axioms in approval-based committee selection relate to one another, with arrows indicating transitive implications. Rectangular boxes represent rules that are difficult to verify, while ellipsoids represent those that admit efficient verification. Solid-line frames denote axioms whose existence is guaranteed or can be checked efficiently, whereas dashed-line frames denote axioms whose existence remains unknown. Lastly, double-line frames indicate axioms for which efficient methods exist to find a solution that satisfies them.
        }
    \label{fig:relation}
\end{figure}

Figure~\ref{fig:relation} provides a visual summary of the current implications among various proportionality axioms, including the new relationships established in this paper.

%% file: fpjr_monroe.tex
\subsection{The Monroe Rule and Greedy Monroe Rule Satisfy FPJR}
Recall that Monroe's Rule aims to find an assignment $\pi: V \rightarrow W$ that assigns each voter to a committee member such that every candidate in the committee is assigned either $\Floor{\frac{n}{k}}$ or $\Ceiling{\frac{n}{k}}$ voters.
For arbitrary coalitions $S$, we define the \emph{Monroe score}
$$M_\pi(S):= \sum_{v \in S} |A_v \cap \SET{\pi(v)}|;$$
the goal is then to find a committee $W$ and assignment $\pi$ maximizing $M_{\pi}(V)$.
The Greedy Monroe Rule approximates this objective by repeatedly adding a candidate $c$ approved by the largest number of unassigned voters, then assigning (at most $\Ceiling{\frac{n}{k}}$) approving voters to $c$, adding arbitrary voters if too few voters approve $c$.

Before delving into the main theme of this section, we present an example illustrating that the Monroe rule does not always satisfy priceability, even when $k \mid n$. This demonstrates that the results in this section are not derivable from those of the previous section.

\begin{example}[The Monroe Rule Violates Priceability]
  \label{ex:monroe-violates-priceability}
  Consider an election with 6 candidates and $n = 6$ voters $V = \{v_1, v_2, v_3, v_4, v_5, v_6\}$, with approval sets as follows:
    \begin{align*}
        A_{v_1} & =\{c_1\} & A_{v_2} & =\{c_2\}
        & A_{v_3} & =A_{v_4} = A_{v_5} = A_{v_6}=\{c_3, c_4, c_5, c_6\}.
    \end{align*}
    The committee size is $k = 3$. The maximum Monroe score is $5$, achieved by selecting one of $c_1$ or $c_2$, along with two candidates from $\{c_3, c_4, c_5, c_6\}$. Without loss of generality, assume that the Monroe rule selects $W = \{c_1, c_3, c_4\}$.

    Suppose that there exists a price system $(p, (p_v)_{v \in V})$ supporting $W$. 
    First, we observe that $p = p_{v_1}(c_1) \leq 1$.
    For voters $v_3, v_4, v_5, v_6$, who only need to support $c_3$ and $c_4$, their total contributions satisfy $\sum_{i=3}^{6} \sum_{c \in C} p_{v_i}(c) \leq 2$. 
    Consequently, the unspent budget across these voters is $4 - \sum_{i=3}^{6} \sum_{c \in C} p_{v_i}(c) \geq 2 > p$. This implies that these voters could collectively afford an additional candidate, such as $c_5$ or $c_6$, contradicting the claim that $(p, (p_v)_{v \in V})$ supports $W$.

    Alternatively, consider $W = \{c_3, c_4, c_5\}$ with $p = \frac{4}{3}$ and each $p_{v_i}(c_j) = \frac{1}{3}$ for $i \in \{3, 4, 5, 6\}$ and $j \in \{3,4,5\}$, and $p_{v_i}(c) = 0$ otherwise. In this case, $(p, (p_v)_{v \in V})$ supports $W$, showing that a priceable committee of size $3$ exists. However, this committee will not be selected by the Monroe rule as it gets Monroe score of $4$. 

    This example highlights that while priceable committees may exist, the Monroe rule does not necessarily select one.
\end{example}

Our main theorem is that both the Monroe Rule and Greedy Monroe Rule satisfy FPJR when $k$ divides $n$.

\begin{theorem} \label{thm:monroe_implies_fpjr}
    When $k$ divides $n$, both the Monroe Rule and the Greedy Monroe Rule satisfy FPJR.
\end{theorem}

For the remainder of this section, we assume that $k$ divides $n$. Before we proceed with the proof, we observe the following about the output of Monroe's Rule. Given an election $(V, C, \A, k)$, let $W$ be the winning committee and $\pi$ an assignment that maximizes the Monroe score.
Even if a cohesive group of voters is not satisfied with the assignment $\pi$, they are nonetheless satisfied with the committee $W$ itself.

\begin{lemma}
    \label{lem:monroe_makes_cohesive_groups_happy}
    If $S \subseteq V$ is a coalition of size $|S| \geq \dfrac{n}{k}$ and $M_\pi(S) = 0$, then $\bigcap_{v \in S} A_v \subseteq W$, i.e., all candidates universally approved by $S$ must be in the committee.
\end{lemma}

\begin{proof}
  Let $c \in \bigcap_{v \in S} A_v$ and assume for contradiction that $c \notin W$.
  Because $M_\pi(S) = 0$, no voter $v \in S$ approves their assigned candidate, i.e., $\pi(v) \notin A_v$ for all voters $v \in S$.
  In fact, for every candidate $c'$ assigned to one or more voter in $S$, we know that $c'$ is not approved by \emph{any} voter in $S$:
  if $\pi(v) = c'$ for some $v \in S$ and $c'$ is approved by $v' \in S$, then by switching the assignments of $v$ and $v'$ (neither of whom approves their currently assigned candidate because $M_\pi(S) = 0$), the Monroe score would strictly increase.
  Now fix such a candidate $c'$ who is assigned to at least one voter in $S$ despite not being approved by any voter in $S$.

  Consider replacing $c'$ with $c$ in the committee, and changing the assignment as follows.
  Let $Y$ be the set of voters not in $S$ who were assigned to $c'$, and let $X \subseteq S$ be a set of voters previously not assigned to $c'$ with $|X| = |Y|$.
  Because $|S| \geq \frac{n}{k}$ and $c'$ has $\frac{n}{k}$ voters assigned to them, such a set $X$ must exist.
  Now define an arbitrary bijection $\phi : Y \to X$, and assign $\pi(\phi(v))$ to each $v \in Y$, and $c$ to each $v \in X$ and each voter $v \in S$ who was previously assigned to $c'$ (leaving all other assignments unchanged).

  Then, while it is possible that no $v \in Y$ approves of their new assigned candidate, all $v \in X$ (of whom there are exactly as many as in $Y$) now go from disapproving to approving their assigned candidate, and the same is true for the candidates in $S$ who were previously assigned to $c'$ (of which there is at least one). Thus, the overall Monroe score strictly increases, contradicting the optimality of the original assignment $\pi$.
\end{proof}

We are now prepared to prove the main theorem.

\begin{proof}[Proof of Theorem~\ref{thm:monroe_implies_fpjr}]
  The proofs for both rules are nearly identical.
  We will therefore combine the proofs, and explicitly point out the parts that are specific to one algorithm or the other.
  
  Let $W$ be the winning set under the selection rule (Monroe or Greedy Monroe).
  Let $S \subseteq V$ be an arbitrary weakly $\ell$-cohesive coalition with witness set $T$, so $\frac{|S|}{n} \geq \frac{|T|}{k}$.
  Assume for contradiction that $S$ violates FPJR, so $W_{S}:= W \cap \bigcup_{v\in S} A_{v}$ has size less than $\ell$.
  
  Now define $S' = S \setminus \bigcup_{c \in W_S} \pi^{-1}(c)$, and let $T' \subseteq T \setminus W_S$ be an arbitrary subset\footnote{Note that $W_S$ is not necessarily a subset of $T$, so $T'$ has to be a strict subset of $T$.} of size $|T| - |W_S|>0$. 

  We first establish that $S'$ is a weakly $(\ell-|W_{S}|)$-cohesive coalition with witness set $T'$.
  Because $|\bigcup_{c \in W_S} \pi^{-1}(c)| \leq |W_S| \cdot \frac{n}{k}$, we get that $|S'| \geq (|T| - |W_S|) \cdot \frac{n}{k} = |T'| \cdot \frac{n}{k}$.
  Thus, $S'$ satisfies the size constraint for being weakly $(\ell-|W_{S}|)$-cohesive.
  Also, because at most $|W_S|$ candidates were removed from $T$ to obtain $T'$, we have that $|A_v \cap T'| \geq |A_v \cap T| - |W_S| \geq \ell - |W_S|$ for all $v \in S'$.
  Thus, $S'$ is indeed weakly $(\ell-|W_{S}|)$-cohesive.

  Next, by definition of $S'$, we know that $\pi(v) \notin W_S$ for all $v \in S'$.
  Because $\pi(v) \in W$ by definition, we obtain that $\pi(v) \notin \bigcup_{v' \in S} A_{v'}$, so in particular, $v$ does not approve of $\pi(v)$.
  This also implies that $M_{\pi}(S') = 0$.

  Because $S'$ is weakly $(\ell-|W_{S}|)$-cohesive, by Lemma~\ref{lem:weak_cohesive_implies_cohesive}, there exists a candidate $c \in T'$ such that $|N_c \cap S'| \geq (\ell - |W_S|) \cdot \frac{n}{k} \geq \frac{n}{k}$.
  Because $|N_c \cap S'| > 0$, $c$ is approved by at least one voter in $S$.
  And because $c \in T'$, we know that $c \notin W_S$, implying that $c \notin W$.
  Now, for the final step of the proof, we distinguish between the two rules:

  \begin{itemize}
    \item If $W$ was the output of the Monroe rule, then we apply Lemma~\ref{lem:monroe_makes_cohesive_groups_happy} to $N_c \cap S'$, and conclude that $c \in W$.
      But this contradicts the earlier conclusion that $c \notin W$.
    \item If $W$ was the output of the Greedy Monroe rule, then consider the first iteration that assigned some voter in $S'$ to a candidate $c'$.
      At that point of the algorithm, $c$ could have been chosen as the next candidate and assigned to $\frac{n}{k}$ voters in $N_c \cap S'$ who all approve $c$, and none of whom had been assigned.
    Instead, the Greedy Monroe rule chose some candidate $c'$ and assigned $c'$ to at least one voter $v$ in $S'$; as stated above, $v$ did not approve $c'$.
    This contradicts the greedy choice of always adding a candidate approved by the largest number of not-yet-assigned voters.
  \end{itemize}
  Thus, we obtained a contradiction in either case, completing the proof of the theorem.
\end{proof}

Together with this theorem and Example~\ref{ex:monroe-violates-priceability} we conclude the section with the following corollary.

\begin{corollary}
    A committee satisfying FPJR is not necessarily priceable.
\end{corollary}

%% file: hardness.tex
\section{Hardness}

In this section, we establish the computational hardness (specifically, coNP-hardness) of verifying FPJR for a proposed committee $W$. Because a very similar reduction also proves hardness for verifying FJR and core stability, and these hardness results seem to not have appeared in the literature, we also fill these gaps.

To prove coNP-hardness of verifying FPJR, we use the exact same reduction which \citet{aziz:elkind:complexity} used to show hardness of verifying PJR; to prove hardness of verifying FJR or core stability, we use the reduction which \citet{aziz:brill:justified} used to show hardness of verifying EJR. However, our proofs are somewhat more involved.
All of the reductions are from the following \textsc{Balanced Biclique} problem:

\begin{definition} \label{def:biclique}
    The \textsc{Balanced Biclique} problem is defined as follows: Given a bipartite graph $G = (L,R)$ and a positive integer $\ell$, determine whether there exist subsets $L' \subseteq L$ and $R'\subseteq R$, each containing $\ell$ vertices, such that all possible edges between $L'$ and $R'$ are present in $E$; that is, $\Set{(u,v)}{u \in L', v \in R'} \subseteq E$. Such a pair $(L', R')$ is called an \emph{$\ell \times \ell$ biclique}.
\end{definition}

The \textsc{Balanced Biclique} problem is known to be NP-complete \citep{garey:johnson:balanced}. We will show that verifying FJR, FPJR, and core stability are coNP-complete by providing reductions from \textsc{Balanced Biclique}.

As mentioned above, our approach closely follows the technique used to establish the hardness of verifying EJR and PJR \citep{aziz:brill:justified, aziz:elkind:complexity}. Specifically, the authors construct a voting instance and a candidate winning set based on an arbitrary instance of \textsc{Balanced Biclique}. They then demonstrate that checking whether EJR or PJR is violated corresponds directly to determining the existence of an $\ell \times \ell$ biclique. In this work, we adopt their constructions and apply more refined reduction arguments.

    In the rest of this section, we provide proofs demonstrating the coNP-completeness of verifying FPJR, FJR, and Core Stability, respectively.

\subsection{Verifying FPJR is coNP-complete}
\begin{theorem}
  The following problem is coNP-complete: Given an arbitrary ballot profile $\A$ and a winning set $W$, does $W$ satisfy FPJR?
\end{theorem}

\begin{algorithm}[h]
    \caption{Construction for Hardness of Verifying PJR \citep{aziz:elkind:complexity}}\label{alg:hardness_pjr}
    \begin{algorithmic}[1]
        \STATE \textbf{Input:} A \textsc{Balanced Biclique} instance $G = (L, R, E)$ and an integer $\ell$, with $|R| \geq \ell \geq 3$.
        \STATE \textbf{Output:} An election $(V, C, \A, k)$ and a designated winner set $W$.
        
        \STATE Let $s \gets |R|$.
        \STATE Define the candidate set $C$ as the union of three sets $C_1, C_2, C_3$ as follows:
            \begin{itemize}
                \item $C_1 = L$,
                \item $|C_2| = \ell - 1$,
                \item $|C_3| = \ell s + 2\ell - 3s - 2$. 
            \end{itemize}
        \STATE Define the voter set $V$ as the union of three sets $V_1, V_2, V_3$ as follows:
            \begin{itemize}
                \item $V_1 = R$,
                \item $|V_2| = \ell \cdot s$,
                \item $|V_3| = \ell s + 2\ell - 3s - 2$.
            \end{itemize}
        \STATE Define an arbitrary bijection $\phi: V_3 \rightarrow C_3$.
        \STATE Define the approval sets $\A$ for each voter $v \in V$ as follows:
        \[
            A_v = 
            \begin{cases}
                \Set{ u \in C_1}{(u, v) \in E} & \text{if } v \in V_1, \\
                C_1 \cup C_2 & \text{if } v \in V_2, \\
                \SET{ \phi(v) } & \text{if } v \in V_3.
            \end{cases}
        \]
        \STATE Set the committee size $k \gets 2 \cdot (\ell - 1)$.
        \STATE Select an arbitrary subset $X \subseteq C_3$ of size $\ell - 1$ and define the winner set $W \gets X \cup C_2$.
        \STATE Return $(V, C, \A, k)$ and $W$.
    \end{algorithmic}
\end{algorithm}

\begin{proof}
  It is easy to see that the problem is in coNP.
  A set of candidates $T \subseteq C$ and a set of voters $S \subseteq V$ which is weakly $\ell$-cohesive with witness set $T$ and $|W \cap \bigcup_{v\in S} A_{v}|<\ell$ gives a certificate for showing that $W$ violates FPJR.

  For coNP-hardness, we again reduce from the \textsc{Balanced Biclique} problem.
  Given an instance $((L, R, E), \ell)$, we construct an election $(V, C, \A, k)$ and a winning set $W$ using Algorithm~\ref{alg:hardness_pjr}, due to \citet{aziz:elkind:complexity}.
  In this construction, notice that
  \begin{center}
      $\frac{|V|}{k} = \frac{s + \ell s + (\ell s + 2\ell - 3s - 2)}{2 (\ell-1)} = s+1$.
  \end{center}
    
  Suppose that there exists an $\ell\times\ell$ biclique $(L', R')$ in $G$.
  Then, as shown by \citep{aziz:elkind:complexity}, PJR would be violated by $R' \cup V_2$.
  Because FPJR implies PJR, this means that FPJR is also violated.

  For the converse direction, assume that FPJR is violated.
  Let $S \subseteq V$ be a weakly $\ell'$-cohesive coalition with witness set $T$ such that $|\bigcup A_v \cap W| < \ell'$.

    We show that $S \subseteq V_1 \cup V_2$.
    For if $S \cap V_3 \neq \emptyset$, then because each voter in $V_3$ approves only of one distinct candidate, we would have $\ell' = 1$, implying that $W \cap A_v = \emptyset$ for all $v \in S$.
    Therefore, $S \subseteq V_{1}\cup V_{3}$, as every voter in $V_2$ approves the winners from $C_2$.
    Because each voter in $V_{3}$ approves a distinct candidate, that candidate must be in $T$, and $|T| \geq |S \cap V_{3}| \geq |S| - |V_{1}| = |S| - s$, or $|S| \leq |T| + s$.
    However, by definition of weak cohesiveness, $|S| \geq \frac{n}{k} \cdot |T| = (s+1) \cdot |T|$, so $|T| = 1$.
    Because there exists a voter $v \in S \cap V_{3}$, $T$ must equal $\SET{\phi(v)}$, but $\phi(v)$ is not approved by any voters in $S \setminus \SET{v}$, thus contradicting the weak cohesiveness of $S$.

    Having established that $S \subseteq V_1 \cup V_2$, we proceed to show that $S \cup T$ contains an $\ell \times \ell$ biclique in the given graph.
    
  Since $|V_1| = s < s + 1 \leq |S|$, $S$ must include at least one voter $v$ from $V_2$.
  This voter $v$ approves all of $C_2 \subseteq W$, so $|A_v \cap W| \geq \ell-1$.
  Because $W$ violates FPJR for $S$, which is weakly $\ell'$-cohesive, we obtain that $\ell'\geq |A_{v}\cap W|\geq \ell-1$, or $\ell' \geq \ell$.
  Because of the size requirement for weak cohesiveness, 
  $|S| \geq \frac{|V|}{k} \cdot |T| = (s+1) \cdot |T|$.
  But $|S| \leq |V_1| + |V_2| = s \cdot (\ell+1)$, implying that $(s+1) \cdot |T| \leq s \cdot (\ell+1)$.
  Hence, $|T| \leq \frac{s}{s+1} \cdot (\ell+1)$, implying that $|T| \leq \ell$.
  But we can also lower-bound the size of $T$ as $|T| \geq |T \cap A_v| \geq \ell'$.
  Therefore, we obtain that $\ell' \leq |T| \leq \ell \leq \ell'$, implying that $|T| = \ell' = \ell$.

  Because each voter $v \in S$ approves of at least $\ell'$ candidates in $T$, each voter in $S$ must approve all candidates in $T$. 
  Moreover, $|S \cap V_1| \geq \ell \cdot (s+1)-|V_{2}| = \ell \cdot (s+1)-\ell s = \ell$, so $S$ contains at least $\ell$ voters from $V_1$. Now let $R'$ be an arbitrary subset of $\ell$ voters from $S \cap V_1$, and $L' = T$.
  We have argued that $(L', R')$ is an $\ell \times \ell$ biclique.
  
  This completes the proof of correctness, and therefore shows that deciding whether $W$ satisfies FPJR is coNP-complete.
\end{proof}

\subsection{Verifying FJR is coNP-complete}
\label{sec:hardness-FJR}
\begin{algorithm}[h]
    \caption{Reduction for Hardness of Verifying EJR \citep{aziz:brill:justified}}\label{alg:hardness_ejr}
    \begin{algorithmic}[1]
        \STATE \textbf{Input:} A \textsc{Balanced Biclique} instance $G = (L, R, E)$ and an integer $\ell$, with $|R| \geq \ell \geq 3$.
        \STATE \textbf{Output:} An election $(V, C, \A, k)$ and a winner set $W$.
        
        \STATE Let $s \gets |R|$.
        \STATE Define the candidate set $C$ as the union of four sets $C_1, C_2, C_3, C_4$ as follows:
            \begin{itemize}
                \item $C_1 = L$,
                \item $|C_2| = |C_3| = \ell - 1$,
                \item $|C_4| = s\ell - 3s + \ell$.
            \end{itemize}
        \STATE {Define the voter set $V$ as the union of three sets $V_1, V_2, V_3$ (of total size $|V|=2s(\ell - 1)$) as follows:}
            \begin{itemize}
                \item $V_1 = R$,
                \item $|V_2| = \ell \cdot (s - 1)$,
                \item $|V_3| = s\ell - 3s + \ell$.
            \end{itemize}
        \STATE Define an arbitrary bijection $\phi: V_3 \rightarrow C_4$.
        \STATE Define the approval sets $\A$ for each voter $v \in V$ as follows:
        \[
            A_v = 
            \begin{cases}
                \{ u \in C_1, (u, v) \in E\} \cup C_2 & \text{if } v \in V_1, \\
                C_1 \cup C_3 & \text{if } v \in V_2, \\
                \{ \phi(v) \} & \text{if } v \in V_3.
            \end{cases}
        \]
        \STATE Set the committee size $k \gets 2 \cdot (\ell - 1)$ and define the winner set $W \gets C_2 \cup C_3$.
        \STATE Return $(V, C, \A, k)$ and $W$.
    \end{algorithmic}
\end{algorithm}

\begin{theorem} \label{thm:FJR-coNP-complete}
    The following problem is coNP-complete: Given an arbitrary ballot profile $\A$ and a winning set $W$, does $W$ satisfy FJR?
\end{theorem}

\begin{proof}
   Membership in coNP is straightforward.
   A certificate for violation of FJR comprises a weakly $\ell$-cohesive set $S \subseteq V$ with witness $T \subseteq C$ such that for each $v \in S$, $|A_v \cap W| < \ell$. This certificate can be verified in polynomial time.
    
    To show coNP-hardness, we reduce from the \textsc{Balanced Biclique} problem. Given an instance $((L, R, E), \ell)$, in polynomial time, construct an election $(V, C, \A, k)$ and a winning set $W$ according to Algorithm~\ref{alg:hardness_ejr} (due to \citet{aziz:brill:justified}). We now verify correctness of the reduction.
    In the following proof, we repeatedly use that $\frac{|V|}{k} = \frac{2\ell s - 2 s}{2(\ell-1)} = s$.
    
    First, suppose that there exists an $\ell \times \ell$ biclique $(L', R')$ in $G$.
        Then, as shown by \citep{aziz:brill:justified}, EJR would be violated by $R' \cup V_2$. Because FJR implies EJR, this means that FJR is also violated.
    
    For the converse direction, assume that $W$ violates FJR, implying the existence of a weakly $\ell'$-cohesive set $S \subseteq V$ with some witness $T$ such that for each $v \in S$, we have $|A_v \cap W| < \ell'$.    
    
    We first show that $S \subseteq V_1 \cup V_2$.
    For if $S \subseteq V_3$, then because each voter in $V_3$ approves a unique candidate, we would obtain that $|T| \geq |S|$, which contradicts $|S| \geq \frac{|V|}{k} \cdot |T| = s \cdot |T| > |T|$.
    If $S$ intersects both $V_3$ and $V_1 \cup V_2$, then because voters in $V_3$ approve exactly one candidate, we must have $\ell'=1$.
    But $S$ contains at least one voter $v \in V_1 \cup V_2$ who approves $\ell-1 \geq 2 > \ell'$ candidates in $W$, so FJR is not violated for $S$.

    We have shown that $S \subseteq V_1 \cup V_2$.
    For each $v \in S$, because FJR is violated, $\ell' > |A_v \cap W| \geq \ell - 1$, implying that $\ell' \geq \ell$.
      Because each voter $v$ approves at least $\ell'$ candidates in $T$, we in particular obtain that $|T| \geq \ell'$.
    On the other hand, $|T| \leq \frac{k}{|V|} \cdot |S| \leq \frac{1}{s} \cdot (|V_1| + |V_2|) = \ell+1 - \frac{\ell}{s}$, so we obtain that $|T| \leq \ell$, and thus $|T| = \ell' = \ell$.

        Because $|T| = \ell = \ell'$, and each voter $v \in S$ approves $\ell'=\ell$ candidates from $T$, we must have that each voter $v$ approves all candidates in $T$.
    Let $R' = S \cap V_1$ and $L' = T$. Since $|S|\geq \ell\cdot s$ and $|V_{2}| = \ell\cdot (s-1)$, we must have $|R'| \geq \ell$, and so $(L', R')$ is an $\ell \times \ell$ biclique in $G$.

    This proves the correctness of the reduction, and thus that verifying FJR is coNP-hard.
\end{proof}

\subsection{Verifying Core Stability is coNP-complete}
\label{sec:hardness-CS}
\begin{theorem} \label{thm:core-stability-coNP-complete}
    The following problem is coNP-hard: Given an arbitrary ballot profile $\A$ and a winning set $W$, is $W$ core-stable for $\A$?
\end{theorem}

\begin{proof}
  Again, membership in coNP is straightforward. A certificate consists of a subset of voters $S \subseteq V$ and a set of candidates $T \subseteq C$ such that $|S| \geq |T| \cdot \frac{|V|}{k}$ and $|A_v \cap T| > |A_v \cap W|$ for each $v \in S$.
  This certificate can be verified in polynomial time, demonstrating that the problem is in coNP.
    
  For coNP-hardness, we again reduce from the \textsc{Balanced Biclique} problem, with the exact same reduction (Algorithm~\ref{alg:hardness_ejr}) as in the proof of Theorem~\ref{thm:FJR-coNP-complete}.
  Let $((L, R, E), \ell)$ be the instance of \textsc{Balanced Biclique}, and $(V, C, \A, k)$ the election and $W$ the winning set produced by Algorithm~\ref{alg:hardness_ejr}.
  Again, recall that under the construction, $\frac{|V|}{k} = s$.
    
    If there exists an $\ell \times \ell$ biclique $(L', R')$ in the input graph, then $W$ violates EJR (as shown by \citep{aziz:brill:justified}), which implies that $W$ cannot be in the core, because core stability implies EJR.

    For the converse direction, assume that $W$ is not in the core.
    Then there exists a set of voters $S \subseteq V$ and a set of candidates $T \subseteq C$ such that $|S| \geq |T| \cdot s$ and for each $v \in S$, we have $|A_v \cap W| < |A_v \cap T|$.
    Without loss of generality, we can assume that 
    $T \subseteq \bigcup_{v \in S} A_v$. Otherwise, one could remove from $T$ all candidates not approved by any voter in $S$, and the resulting set $T$ would still be a (smaller) preferred deviation for $S$.

    Next, we show that without loss of generality, we can assume that $S \subseteq V_1 \cup V_2$.
    Indeed, consider $S' = S \setminus V_3$ and $T' = T \setminus C_4$.
    Because each voter in $V_3 \cap S$ approves only a single unique candidate in $C_4$, and must approve at least one candidate in $T$, we obtain that $|T \cap C_4| \geq |S \cap V_3|$, and hence
    \[
    \frac{|S'|}{|T'|} = \frac{|S| - |S \cap V_3|}{|T| - |T \cap C_4|} \geq \frac{|S| - |S \cap V_3|}{|T| - |S \cap V_3|} \geq \frac{|S|}{|T|} \geq s.
    \]
    In addition, because no voter in $v \in V_1 \cup V_2$ approves any candidate in $C_4$, we have that $|A_v \cap T'| = |A_v \cap T| > |A_v \cap W|$ for all $v \in S'$.
    This shows that if $T$ is a deviation for $S$, then $T'$ is a deviation for $S'$.
    Hence, we assume from now on that $S \subseteq V_1 \cup V_2$.
    
    Now, we will show that $S \cup T$ contains an $\ell \times \ell$ biclique in $G$.
    Since for each $v \in S$, we have that $|A_v \cap T| > |A_v \cap W| = \ell - 1$, it follows that $|T| \geq \ell$.
    On the other hand, $|T| \leq \frac{k}{|V|} \cdot |S| \leq \frac{1}{s} \cdot (|V_1| + |V_2|) = \ell+1 - \frac{\ell}{s} < \ell+1$, so $|T| = \ell$.
    Hence, all voters in $S$ must approve all candidates in $T$. This is exactly the situation from the final paragraph of the proof of Theorem~\ref{thm:FJR-coNP-complete}, and as in that proof, we conclude that $L'=T$ and $R' \subseteq S \cap V_1$ of size $\ell$ defines an $\ell \times \ell$ biclique.
    
    This shows correctness of the reduction, and hence coNP-completeness of verifying core stability.
    \end{proof}

%% file: conclusion.tex
\section{Conclusion}
\label{sec:conclusion}

We presented a new proportionality axiom called Full Proportional Justified Representation (FPJR), situated between PJR and FJR, but incomparable to EJR.
We related FPJR to various other notions of proportionality, and showed that several well-known algorithms output committees satisfying FPJR.
As for PJR, EJR, and FJR, verifying whether a given committee satisfies FPJR is coNP-complete.

Our notion grew out of a desire to understand FJR better.
Finding an efficient algorithm for computing a committee satisfying FJR is still one of the central open questions in the area, along with understanding if the core is always guaranteed to be non-empty.

FPJR, similar to EJR, PJR, and FJR, can be naturally extended to participatory budgeting contexts.
Further exploration of the axiomatic foundations and algorithmic properties of FPJR in such scenarios offers an interesting avenue for future research.

There is increasing interest in proportionality axioms that allow polynomial-time verification, which we showed FPJR to not allow.
An intriguing open question is whether stronger axioms than FPJR or FJR can be defined while retaining efficient verifiability, akin to the enhanced axioms EJR+ and PJR+.

Finally, we demonstrated that priceability is a stricter concept than FPJR: some committees satisfy FPJR but not priceability.

%% file: appendix.tex
\section{Relation between FPJR and PJR+}

\subsection{FPJR and PJR+ Are Incomparable}
\label{apx:fpjr_vs_pjr+}
As we discussed earlier, the output of the PAV rule satisfies PJR+ but may violate FPJR.
Here, we present an example where FPJR is satisfied but PJR+ is violated, demonstrating that FPJR and PJR+ are incomparable.

\begin{example}
  Consider an election $(V, C, \A, k)$ with $|V| = 12$, $C = \SET{c_1, c_2, c_3, c_4, c_5, c_6, c_7}$, and approval ballots $\A$ defined as: 
    \begin{itemize} 
        \item Voters $v_1, v_2, v_3$ approve $\SET{c_1, c_2}$. 
        \item Voters $v_4, v_5, v_6$ approve $\SET{c_1, c_3}$. 
        \item Voters $v_7, v_8, v_9, v_{10}, v_{11}, v_{12}$ approve $\SET{c_4, c_5, c_6, c_7}$. 
    \end{itemize} 
    Let $k = 6$, and note that $\frac{n}{k} = 2$.
    Consider the winning set $W = \SET{c_2, c_3, c_4, c_5, c_6, c_7}$.
    Any coalition of voters contained entirely in $\SET{v_1, v_2, v_3}$ or in $\SET{v_4, v_5, v_6}$ is too small to be more than weakly 1-cohesive and approves one candidate from $W$. Any coalition including voters from both sets is at most weakly 2-cohesive, and jointly approves at least two candidates in $W$. And any coalition including voters $v_i$ with $i \geq 7$ is at most 4-cohesive and approves four candidates in $W$. Thus, $W$ satisfies FPJR.

    However, candidate $c_1$ is approved by all voters in $S = \SET{v_1, v_2, v_3, v_4, v_5, v_6}$ and not elected, so $S$ is a $3$-deprived set with utility 2, violating PJR+. 
\end{example}

%% file: main.bbl
\begin{thebibliography}{35}
\providecommand{\natexlab}[1]{#1}
\providecommand{\url}[1]{\texttt{#1}}
\expandafter\ifx\csname urlstyle\endcsname\relax
  \providecommand{\doi}[1]{doi: #1}\else
  \providecommand{\doi}{doi: \begingroup \urlstyle{rm}\Url}\fi

\bibitem[Aziz and Lee(2021)]{aziz:lee:proportional-pb-ordinal}
Haris Aziz and Barton~E. Lee.
\newblock Proportionally representative participatory budgeting with ordinal
  preferences.
\newblock In \emph{Proc. 35th AAAI Conf. on Artificial Intelligence}, pages
  5110--5118, 2021.
\newblock \doi{10.1609/aaai.v35i6.16646}.

\bibitem[Aziz and Shah(2021)]{aziz:shah:pb}
Haris Aziz and Nisarg Shah.
\newblock Participatory budgeting: Models and approaches.
\newblock In Tam{\'a}s Rudas and G{\'a}bor P{\'e}li, editors, \emph{Pathways
  Between Social Science and Computational Social Science: Theories, Methods,
  and Interpretations}, pages 215--236. Springer International Publishing,
  2021.
\newblock ISBN 978-3-030-54936-7.
\newblock \doi{10.1007/978-3-030-54936-7_10}.
\newblock URL \url{https://doi.org/10.1007/978-3-030-54936-7_10}.

\bibitem[Aziz et~al.(2014)Aziz, Gaspers, Gudmundsson, Mackenzie, Mattei, and
  Walsh]{aziz:gaspers:computational}
Haris Aziz, Serge Gaspers, Joachim Gudmundsson, Simon Mackenzie, Nicholas
  Mattei, and Toby Walsh.
\newblock Computational aspects of multi-winner approval voting.
\newblock In \emph{Workshops at the Twenty-Eighth AAAI Conference on Artificial
  Intelligence}, 2014.

\bibitem[Aziz et~al.(2017)Aziz, Brill, Conitzer, Elkind, Freeman, and
  Walsh]{aziz:brill:justified}
Haris Aziz, Markus Brill, Vincent Conitzer, Edith Elkind, Rupert Freeman, and
  Toby Walsh.
\newblock Justified representation in approval-based committee voting.
\newblock \emph{Social Choice and Welfare}, 48\penalty0 (2):\penalty0 461--485,
  2017.

\bibitem[Aziz et~al.(2018{\natexlab{a}})Aziz, Elkind, Huang, Lackner,
  S{\'a}nchez-Fern{\'a}ndez, and Skowron]{aziz:elkind:complexity}
Haris Aziz, Edith Elkind, Shenwei Huang, Martin Lackner, Luis
  S{\'a}nchez-Fern{\'a}ndez, and Piotr Skowron.
\newblock On the complexity of extended and proportional justified
  representation.
\newblock In \emph{Proc. 32nd AAAI Conf. on Artificial Intelligence}, pages
  902--909, 2018{\natexlab{a}}.

\bibitem[Aziz et~al.(2018{\natexlab{b}})Aziz, Lee, and
  Talmon]{aziz:lee:talmon:pb}
Haris Aziz, Barton~E. Lee, and Nimrod Talmon.
\newblock Proportionally representative participatory budgeting: Axioms and
  algorithms.
\newblock In \emph{Proc. 17th Intl. Conf. on Autonomous Agents and Multiagent
  Systems}, pages 23--31. International Foundation for Autonomous Agents and
  Multiagent Systems, 2018{\natexlab{b}}.

\bibitem[Brams et~al.(2022)Brams, Brill, and George]{brams:brill:excess-method}
Steven~J. Brams, Markus Brill, and Anne-Marie George.
\newblock The excess method: a multiwinner approval voting procedure to
  allocate wasted votes.
\newblock \emph{Social Choice and Welfare}, 58\penalty0 (2):\penalty0 283--300,
  February 2022.
\newblock ISSN 1432-217X.
\newblock \doi{10.1007/s00355-021-01358-3}.
\newblock URL \url{https://doi.org/10.1007/s00355-021-01358-3}.

\bibitem[Brandt et~al.(2016)Brandt, Conitzer, Endriss, Lang, and
  Procaccia]{brandt:conitzer:handbook}
Felix Brandt, Vincent Conitzer, Ulle Endriss, J\'{e}r\^{o}me Lang, and Ariel~D.
  Procaccia.
\newblock \emph{Handbook of Computational Social Choice}.
\newblock Cambridge University Press, USA, 1st edition, 2016.
\newblock ISBN 1107060435.

\bibitem[Brill and Peters(2023)]{brill:peters:verifiable-JR}
Markus Brill and Jannik Peters.
\newblock Robust and verifiable proportionality axioms for multiwinner voting.
\newblock In \emph{Proc. 24th ACM Conf. on Economics and Computation}, page
  301. Association for Computing Machinery, 2023.
\newblock ISBN 9798400701047.
\newblock \doi{10.1145/3580507.3597785}.
\newblock URL \url{https://doi.org/10.1145/3580507.3597785}.

\bibitem[Brill et~al.(2023)Brill, Forster, Lackner, Maly, and
  Peters]{brill:forster:abc-pb-proportionality}
Markus Brill, Stefan Forster, Martin Lackner, Jan Maly, and Jannik Peters.
\newblock Proportionality in approval-based participatory budgeting.
\newblock In \emph{Proc. 37th AAAI Conf. on Artificial Intelligence}. AAAI
  Press, 2023.
\newblock ISBN 978-1-57735-880-0.
\newblock \doi{10.1609/aaai.v37i5.25686}.
\newblock URL \url{https://doi.org/10.1609/aaai.v37i5.25686}.

\bibitem[Brill et~al.(2024)Brill, Freeman, Janson, and
  Lackner]{brill:freeman:phragmen-JR}
Markus Brill, Rupert Freeman, Svante Janson, and Martin Lackner.
\newblock Phragm\'{e}n's voting methods and justified representation.
\newblock \emph{Mathematical Programming}, 203\penalty0 (1):\penalty0 47--76,
  2024.
\newblock ISSN 1436-4646.
\newblock \doi{10.1007/s10107-023-01926-8}.
\newblock URL \url{https://doi.org/10.1007/s10107-023-01926-8}.

\bibitem[Caragiannis et~al.(2022)Caragiannis, Shah, and
  Voudouris]{caragiannis:shah:metric-distortion}
Ioannis Caragiannis, Nisarg Shah, and Alexandros~A. Voudouris.
\newblock The metric distortion of multiwinner voting.
\newblock In \emph{Proc. 36th AAAI Conf. on Artificial Intelligence}, pages
  4900--4907, 2022.
\newblock \doi{10.1609/aaai.v36i5.20419}.

\bibitem[Chamberlin and Courant(1983)]{chamberlin:courant}
John~R. Chamberlin and Paul~N. Courant.
\newblock Representative deliberations and representative decisions:
  Proportional representation and the borda rule.
\newblock \emph{American Political Science Review}, 77\penalty0 (3):\penalty0
  718–733, 1983.
\newblock \doi{10.2307/1957270}.

\bibitem[Dummett(1984)]{dummett-voting-procedures}
Michael Dummett.
\newblock \emph{Voting Procedures}.
\newblock Oxford University Press, 1984.

\bibitem[Ebadian and Micha(2024)]{ebadian:micha:boosting-sortition}
Soroush Ebadian and Evi Micha.
\newblock Boosting sortition via proportional representation, 2024.
\newblock URL \url{https://arxiv.org/abs/2406.00913}.

\bibitem[Faliszewski et~al.(2017)Faliszewski, Skowron, Slinko, and
  Talmon]{faliszewski:skowron:multiwinner}
Piotr Faliszewski, Piotr Skowron, Arkadii Slinko, and Nimrod Talmon.
\newblock Multiwinner voting: {A} new challenge for social choice theory.
\newblock \emph{Trends in computational social choice}, 74\penalty0
  (2017):\penalty0 27--47, 2017.
\newblock Publisher: AI Access Foundation El Segundo.

\bibitem[Faliszewski et~al.(2018)Faliszewski, Slinko, Stahl, and
  Talmon]{faliszewski:slinko:greedy-monroe}
Piotr Faliszewski, Arkadii Slinko, Kolja Stahl, and Nimrod Talmon.
\newblock Achieving fully proportional representation by clustering voters.
\newblock \emph{Journal of Heuristics}, 24\penalty0 (5):\penalty0 725–756,
  2018.
\newblock ISSN 1381-1231.
\newblock \doi{10.1007/s10732-018-9376-y}.
\newblock URL \url{https://doi.org/10.1007/s10732-018-9376-y}.

\bibitem[Garey and Johnson(1979)]{garey:johnson:balanced}
Michael~R. Garey and David~S. Johnson.
\newblock \emph{Computers and intractability}.
\newblock Freeman San Francisco, 1979.

\bibitem[Goel et~al.(2019)Goel, Krishnaswamy, Sakshuwong, and
  Aitamurto]{goel:krishnaswamy:knapsack-voting}
Ashish Goel, Anilesh~K. Krishnaswamy, Sukolsak Sakshuwong, and Tanja Aitamurto.
\newblock Knapsack voting for participatory budgeting.
\newblock \emph{ACM Transactions on Economics and Computation}, 7\penalty0 (2),
  July 2019.
\newblock ISSN 2167-8375.
\newblock \doi{10.1145/3340230}.
\newblock URL \url{https://doi.org/10.1145/3340230}.

\bibitem[Humphreys(1911)]{humphreys:proportional-representation}
John~H. Humphreys.
\newblock \emph{Proportional Representation: A Study in Methods of Election}.
\newblock Methuen \& Co, 1911.

\bibitem[Janson(2018)]{janson:phragmen-thiele}
Svante Janson.
\newblock Phragm\'{e}n's and thiele's election methods, 2018.
\newblock URL \url{https://arxiv.org/abs/1611.08826}.

\bibitem[Kalayci et~al.(2024)Kalayci, Kempe, and
  Kher]{kalayci:kempe:kher:proportional-representation-metric}
Yusuf Kalayci, David Kempe, and Vikram Kher.
\newblock Proportional representation in metric spaces and low-distortion
  committee selection.
\newblock \emph{Proc. 38th AAAI Conf. on Artificial Intelligence}, 38:\penalty0
  9815--9823, 2024.
\newblock \doi{10.1609/aaai.v38i9.28841}.
\newblock URL \url{https://ojs.aaai.org/index.php/AAAI/article/view/28841}.

\bibitem[Lackner and Skowron(2023)]{lackner:skowron:abc-survey}
Martin Lackner and Piotr Skowron.
\newblock \emph{Approval-Based Committee Voting}, pages 1--7.
\newblock Springer International Publishing, Cham, 2023.
\newblock ISBN 978-3-031-09016-5.
\newblock \doi{10.1007/978-3-031-09016-5_1}.
\newblock URL \url{https://doi.org/10.1007/978-3-031-09016-5_1}.

\bibitem[Monroe(1995)]{monroe:rule}
Burt~L. Monroe.
\newblock Fully proportional representation.
\newblock \emph{American Political Science Review}, 89\penalty0 (4):\penalty0
  925–940, 1995.
\newblock \doi{10.2307/2082518}.

\bibitem[Moulin(1986)]{moulin:choosing-tournament}
Herv\'{e} Moulin.
\newblock Choosing from a tournament.
\newblock \emph{Social Choice and Welfare}, 3\penalty0 (4):\penalty0 271--291,
  1986.

\bibitem[Peters and Skowron(2020{\natexlab{a}})]{peters:skowron:mes}
Dominik Peters and Piotr Skowron.
\newblock Proportionality and the limits of welfarism.
\newblock In \emph{Proc. 21st ACM Conf. on Economics and Computation}, page
  793–794. Association for Computing Machinery, 2020{\natexlab{a}}.
\newblock ISBN 9781450379755.
\newblock \doi{10.1145/3391403.3399465}.
\newblock URL \url{https://doi.org/10.1145/3391403.3399465}.

\bibitem[Peters and Skowron(2020{\natexlab{b}})]{peters:skowron:priceability}
Dominik Peters and Piotr Skowron.
\newblock Proportionality and the limits of welfarism.
\newblock In \emph{Proc. 21st ACM Conf. on Economics and Computation}, pages
  793--–794, New York, NY, USA, 2020{\natexlab{b}}. Association for Computing
  Machinery.
\newblock ISBN 9781450379755.
\newblock \doi{10.1145/3391403.3399465}.
\newblock URL \url{https://doi.org/10.1145/3391403.3399465}.

\bibitem[Peters et~al.(2021{\natexlab{a}})Peters, Pierczy{\'n}ski, and
  Skowron]{peters:grzegorz:mes-pb}
Dominik Peters, Grzegorz Pierczy{\'n}ski, and Piotr Skowron.
\newblock Proportional participatory budgeting with additive utilities.
\newblock \emph{Proc. 35th Advances in Neural Information Processing Systems},
  34:\penalty0 12726--12737, 2021{\natexlab{a}}.

\bibitem[Peters et~al.(2021{\natexlab{b}})Peters, Pierczy\'{n}ski, and
  Skowron]{peters:pierczynski:skowron:fjr}
Dominik Peters, Grzegorz Pierczy\'{n}ski, and Piotr Skowron.
\newblock Proportional participatory budgeting with additive utilities.
\newblock In \emph{Proc. 35th Advances in Neural Information Processing
  Systems}, Red Hook, NY, USA, 2021{\natexlab{b}}. Curran Associates Inc.
\newblock ISBN 9781713845393.

\bibitem[Pierczy\'{n}ski and Skowron(2019)]{pierczynski:skowron:approval}
Grzegorz Pierczy\'{n}ski and Piotr Skowron.
\newblock Approval-based elections and distortion of voting rules.
\newblock In \emph{Proc. 28th Intl. Joint Conf. on Artificial Intelligence},
  pages 543--549, 2019.

\bibitem[Rey and Maly(2023)]{maly:rey:pbsurvey}
Simon Rey and Jan Maly.
\newblock The (computational) social choice take on indivisible participatory
  budgeting, 2023.
\newblock URL \url{https://arxiv.org/abs/2303.00621}.

\bibitem[S{\'a}nchez-Fern{\'a}ndez et~al.(2017)S{\'a}nchez-Fern{\'a}ndez,
  Elkind, Lackner, Fern{\'a}ndez, Fisteus, Val, and
  Skowron]{sanchez:elkind:pjr}
Luis S{\'a}nchez-Fern{\'a}ndez, Edith Elkind, Martin Lackner, Norberto
  Fern{\'a}ndez, Jes{\'u}s Fisteus, Pablo~Basanta Val, and Piotr Skowron.
\newblock Proportional justified representation.
\newblock In \emph{Proc. 31st AAAI Conf. on Artificial Intelligence}, pages
  670--676, 2017.

\bibitem[S\'{a}nchez-Fern\'{a}ndez et~al.(2024)S\'{a}nchez-Fern\'{a}ndez,
  Fisteus, and L\'{o}pez-Zaragoza]{sanchez:fisteus:approval-data}
Luis S\'{a}nchez-Fern\'{a}ndez, Jes'{u}s~A. Fisteus, and Rafael
  L\'{o}pez-Zaragoza.
\newblock Data as voters: instance selection using approval-based multi-winner
  voting, 2024.
\newblock URL \url{https://arxiv.org/abs/2304.09995}.

\bibitem[Skowron et~al.(2017)Skowron, Lackner, Brill, Peters, and
  Elkind]{skowron:lackner:proportional-rankings}
Piotr Skowron, Martin Lackner, Markus Brill, Dominik Peters, and Edith Elkind.
\newblock Proportional rankings.
\newblock In \emph{Proc. 26th Intl. Joint Conf. on Artificial Intelligence},
  pages 409--415, 2017.
\newblock \doi{10.24963/ijcai.2017/58}.
\newblock URL \url{https://doi.org/10.24963/ijcai.2017/58}.

\bibitem[Thiele(1895)]{thiele:pav}
Thorvald~N. Thiele.
\newblock Om flerfoldsvalg.
\newblock \emph{Oversigt over det Kongelige Danske Videnskabernes Selskabs
  Forhandlinger}, 1895:\penalty0 415--441, 1895.

\end{thebibliography}
